\documentclass[]{aastex631}
\usepackage{epsfig}
\usepackage{graphicx}
\usepackage{longtable}
\usepackage{natbib}
\usepackage{subfigure}
\usepackage{xcolor}
\usepackage{amsmath}
\usepackage{multirow}
\usepackage{todonotes}
\usepackage{subfigure}

\newcommand{\Msolar}{M$_{\odot}$}

\begin{document}
\shorttitle{COS Spectroscopy of a Blue Lurker in M67}
\shortauthors{Leiner et al.}

\title{The Blue Lurker WOCS 14020: A Long-Period Post-Common-Envelope Binary in M67 Originating from a Merger in a Triple System }

\author{Emily M. Leiner}
\affiliation{Department of Physics, Illinois Institute of Technology, 3101 South Dearborn St., Chicago, IL 60616, USA}
\email{eleiner@iit.edu}
\affiliation{Center for Interdisciplinary Exploration and Research in Astrophysics (CIERA), Northwestern University, 1800 Sherman Ave., Evanston, IL 60201, USA}

\author{Natalie M. Gosnell}
\affiliation{Department of Physics, Colorado College, 14 E. Cache La Poudre St., Colorado Springs, CO 80903, USA}

\author{Aaron M. Geller}
\affiliation{Center for Interdisciplinary Exploration and Research in Astrophysics (CIERA) and Department of Physics and Astronomy, Northwestern University, 1800 Sherman Ave., Evanston, IL 60201, USA}

\author{Meng Sun}
\affiliation{Center for Interdisciplinary Exploration and Research in Astrophysics (CIERA), Northwestern University, 1800 Sherman Ave., Evanston, IL 60201, USA}

\author{Robert D. Mathieu}
\affiliation{Department of Astronomy, University of Wisconsin - Madison, 475 N. Charter Street, Madison WI 53726}

\author{Alison Sills}
\affiliation{Department of Physics and Astronomy, McMaster University, 1280 Main Street West, Hamilton, ON, L8S 3X6, CANADA}

\begin{abstract}
We present \textit{Hubble Space Telescope } far-ultraviolet (FUV) spectra of a blue-lurker$-$white-dwarf (BL-WD) binary system in the 4 Gyr open cluster M67. We fit the FUV spectrum of the WD, determining it is a C/O WD with a mass of $0.72^{+0.05}_{-0.04}$ \Msolar\ and a cooling age of $\sim400$ Myr. This requires a WD progenitor of $\sim3$ \Msolar, significantly larger than the current cluster turnoff mass of 1.3~\Msolar. We suggest the WD progenitor star formed several hundred Myr ago via the merger of two stars near the turnoff of the cluster. In this scenario, the original progenitor system was a hierarchical triple consisting of a close, near-equal-mass inner binary, with a tertiary companion with an orbit of a few thousand days. The WD is descended from the merged inner binary, and the original tertiary is now the observed BL. The likely formation scenario involves a common envelope while the WD progenitor is on the AGB, and thus the observed orbital period of 359 days requires an efficient common envelope ejection. The rapid rotation of the BL indicates it accreted some material during its evolution, perhaps via a wind prior to the common envelope. This system will likely undergo a second common envelope in the future, and thus could result in a short-period double WD binary or merger of a 0.72~\Msolar\ C/O WD and a 0.38~\Msolar\ Helium WD, making this a potential progenitor of an interesting transient such as a sub-Chandrasekhar Type Ia supernova. 
\end{abstract}

\section{Introduction}
\begin{figure}
    \centering
    \includegraphics[width= 0.9\linewidth]{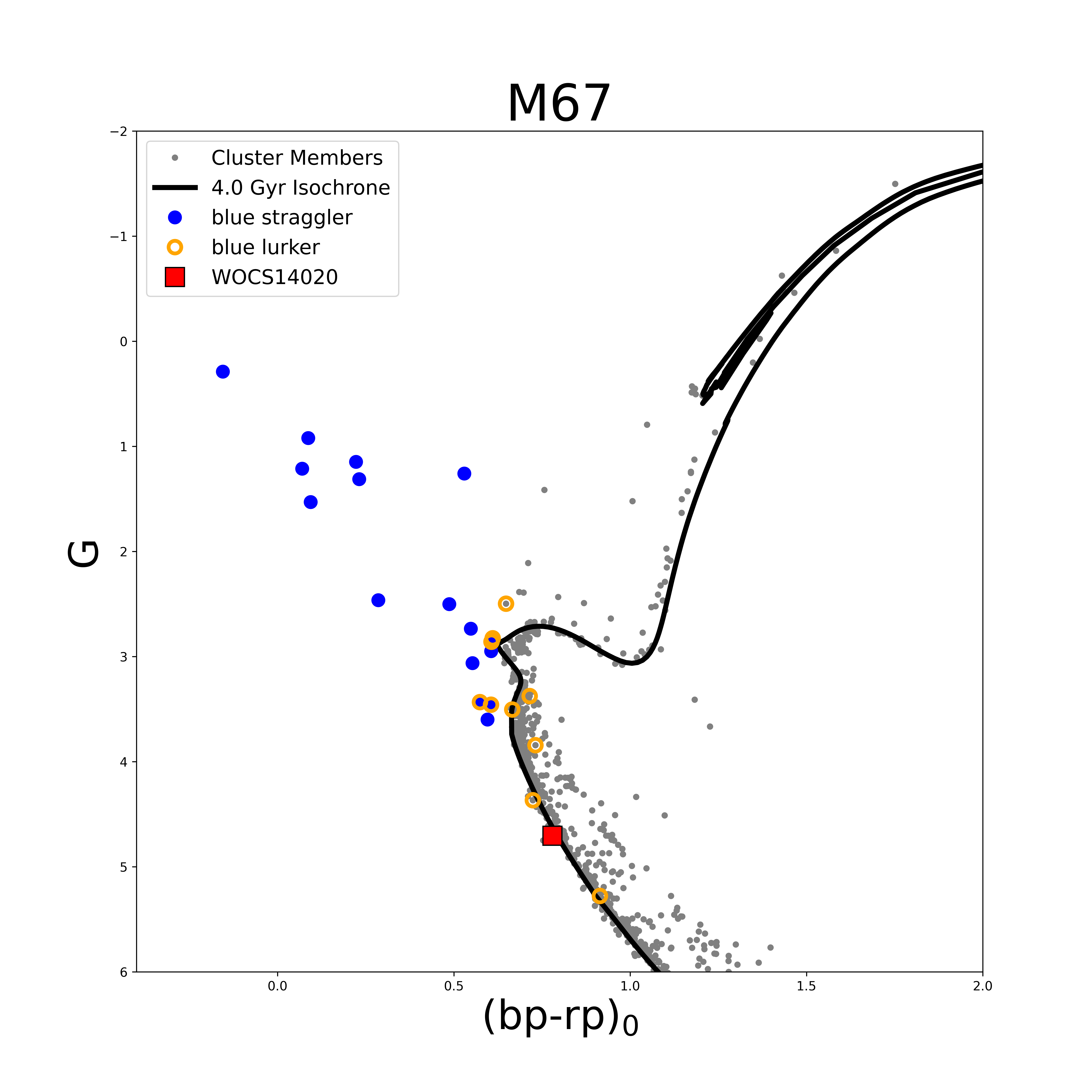}
    \caption{Color-magnitude diagram of M67 highlighting \textit{Gaia} DR3 proper-motion members (gray), blue stragglers (blue), BLs (orange), and our target, the BL WOCS 14020 (red square). Memberships are determined as described in \citet{Leiner2021}}
    \label{fig:CMD}
\end{figure}

Blue lurkers (BLs) are stars that appear in color-magnitude diagrams to be normal main sequence stars (see Figure~\ref{fig:CMD}), but they have shorter rotation periods than expected. For instance, the solar-like stars in the old (4 Gyr) open cluster M67 rotate with $P_{rot} \sim 20-30$ days. \citet{Leiner2019} detected 11 BLs in M67 with $P_\text{rot} \leq 8$ days. To explain this rapid rotation, \citet{Leiner2019} hypothesized BLs have been spun up via stellar mergers, collisions, or mass transfer in binary systems. BLs are therefore thought to be the lower-luminosity counterparts to the more well-known blue straggler stars, which are stars brighter than the main-sequence turnoff found in open and globular clusters (Figure~\ref{fig:CMD}). Like the blue stragglers, BLs may be formed via mass transfer from a giant companion \citep{McCrea1964, Chen2008}, stellar mergers \citep{Perets2009}, or stellar collisions during dynamical encounters \citep{Knigge2009, Leigh2011}.

An observational test of the mass-transfer hypothesis for blue straggler formation is to search for white dwarf (WD) companions to blue straggler stars. Studies of the old (6 Gyr) open cluster NGC 188 used \textit{HST} UV photometry and spectroscopy to identify WD companions to blue stragglers \citep{Gosnell2014, Gosnell2015, Gosnell2019}, finding that $\sim 1/3$ of the blue stragglers had detectable hot WD companions in orbits of $10^2-- 10^3$ days. Further, these studies suggested that $2/3$ of the blue straggler population likely had WD companions, as some WDs would be too cool and old to be detected.  Numerous  UV detections of WD companions to blue stragglers have now been claimed in other clusters using far-UV imaging (e.g. \citealt{Saketh2024, Jadhav2019, Sindhu2019}, \citealt{Panthi2022, Panthi2024} and references therein).  Direct detection of UV flux from WD companions demonstrates definitively that many blue stragglers form via transfer from a giant companion. 

\citet{Nine2023} applied this same technique to the BL population of M67 identified in \citet{Leiner2019}, using \textit{HST} UV photometry to look for UV excesses indicative of WD companions. They detected hot, young ($< 900$ Myr) WDs in two of the BL binary systems (WOCS 3001 and WOCS 14020), confirming that at least $\sim 20 \%$ of the BLs have been spun up to rapid rotation rates via mass transfer from a giant binary companion. The true fraction of mass transfer formation is likely larger as older, cooler WD companions are too faint to be photometrically detected. \citet{Jadhav2019} also report UV excesses to M67 BLs WOCS 3001 and possibly WOCS 9005 using AstroSat/UVIT, though the WOCS 9005 detection was not confirmed by \citet{Nine2023}. The detection of WD companions to some BLs solidifies the hypothesis that BLs are lower-luminosity, lower mass analogs to the blue straggler stars that blend photometrically with typical main sequence stars, and that some formed via  mass transfer from a red giant branch (RGB) or asymptotic giant branch (AGB) donor stars. 

While many blue stragglers, BLs and related systems are now known to have WD companions, and thus to have formed via mass transfer of some kind, very few have the detailed stellar and orbital parameters needed to infer detailed formation histories. WD masses and ages, in particular, are essential constraints because they define the evolutionary state of the donor star at the end of mass transfer, and the time since mass transfer occurred. These parameters can only be ascertained from eclipsing or self lensing binaries (e.g. \citealt{Kawahara2018}) or from fitting FUV WD spectra (e.g. \citep{Landsman1997}) and therefore only a few blue stragglers have well characterized formation histories \citep{Landsman1997, Brogaard2018, Gosnell2019, Sun2021, Sun2023}, and no blue lurkers have yet had a detailed formation history proposed. 

Post-mass-transfer blue stragglers and BLs have orbital periods of $10^2-10^3$ days. These long orbital periods challenge the often used assumption in population synthesis models that mass transfer from more massive giant stars on to less massive main-sequence accretors should be unstable, leading to a common envelope and orbital inspiral \citep{Hjellming1987} to form short-period binaries with orbital periods of just a few days. The wide orbital periods of the BLs may support recent theoretical models that suggest stable mass transfer may occur onto lower-mass accretors than canonically predicted (see, for example \citealt{Temmink2023, Ge2020, Pavlovskii2015, Passy2012, Woods2011}). Alternatively, it may be that BLs are not the result of standard stable mass transfer, but instead form via another pathway such as via wind accretion. Even minimal accretion via a wind ($\Delta M < 0.1$ \Msolar) has been shown to potentially spin accreting stars up to velocities approaching break up \citep{Sun2024}, potentially explaining the observed rapid rotation in BLs. 

In the few cases where the mass-transfer histories of blue stragglers have been investigated in detail, the formation pathways include both quite conservative and highly non-conservative mass transfer on the RGB \citep{Landsman1997, Sun2021}, and a combination of Roche lobe overflow on the AGB and wind mass transfer \citep{Sun2023}. Thus, formation paths of blue stragglers and blue lurkers appear to be varied, and more detailed case studies are needed to better understand the range of formation scenarios and formation physics.

Here we seek to understand the formation pathway of a BL-WD binary in M67, WOCS 14020, by constraining the WD mass and cooling age from its far-UV spectrum in order to re-construct the evolutionary history of the system. In Section~\ref{sec:target} we describe the targeted BL-WD binary, WOCS 14020. In Section~\ref{sec:analysis} we present the observation and spectral analysis technique we use to determined a WD mass and cooling age for this target. These masses and ages inform the possible mass transfer history of this system, which we explore with some modeling in Section~\ref{sec:formation}. We discuss the implications of our findings (Section~\ref{sec:discussion}) and conclude with a summary of our results (Section ~\ref{sec:summary}). 

\section{Target and Observations}
\label{sec:target}
\subsection{WOCS 14020}
WOCS 14020 is a binary star system in the 4 Gyr, solar-metallicity cluster M67 \citep{Geller2015, Leiner2019, Geller2021}. It is a single-lined spectroscopic binary dominated by the light from the BL primary star. Given this system's position in the CMD (Figure~\ref{fig:CMD}), the BL star is consistent with a $\sim1.0$ \Msolar\ main-sequence star. The BL primary was found in \citet{Leiner2019} to have a rotation period of 4.4 days, unusually fast for a solar-like star at an age of 4 Gyr, which have expected rotation periods $> 20$ days. Gyrochronology models \citep{Angus2019} predict an age of $\sim300$ Myr given this rotation period. \citet{Nine2023} analyzed \textit{HST} FUV photometry of WOCS 14020, detecting a FUV excess consistent with a hot WD companion. The photometry is consistent with a temperature for the WD of $\sim11,000-13,000$ K, implying a time since mass-transfer formation of $\sim 290-540$ Myr, consistent with the gyrochonology age. Assuming a primary mass of 1.05 \Msolar, the binary mass function also yields a very low minimum secondary mass (0.15 \Msolar), consistent with a WD secondary. We summarize the system properties of the BL primary and the orbital parameters of the binary system in Table~\ref{tab:WOCS14020}. 


\subsection{HST Spectroscopy}
WOCS 14020 was observed by \textit{HST} Cosmic Origins Spectrograph (COS) for GO program 17134 over five separate visits of two orbits each.  It was observed in TIME-TAG mode through the Primary Science Aperture (PSA) using the G140L grating with a central wavelength of 1105\AA. This region covers the Lyman-$\alpha$ wings and provides the wide wavelength coverage necessary to fit WD atmosphere models in this region. To increase the signal-to-noise of the final spectrum, we co-added the MAST-reduced spectrum for each visit and binned the resulting spectrum to a new wavelength resolution of 3.0\AA. The spectrum is dereddened using \texttt{pysynphot} tools assuming $E(B-V)=0.041$ \citep{Taylor2007}. Geocoronal emission lines are evident in the spectrum and are masked by hand from the analysis. 

\begin{table}[]
    \centering
    \begin{tabular}{c|c|c}
        \hline
        & & Comment \\
        \hline
        \hline
        $M$  & $\sim1.05$ \Msolar\ & Estimated from evolutionary track fit to photometry \\
        $T_\text{eff}$& $5990^{+60}_{-100}$ K & SED fit in \citet{Nine2023}\\
        $M_G$ & 4.76  & Gaia DR3 absolute magnitude using $d= 816$ pc and $E(B-V)= 0.041$\\
        $(bp-rp)_0$  & 0.78 & Gaia DR3 color with E(B-V)= 0.041 and \citet{Wang2019} extinction law\\
        $P_\text{rot}$ &  4.4 days & From \citet{Leiner2019}\\
        $P_\text{orb}$  & 358.9 days & From \citet{Leiner2019}\\
        $ecc$   & 0.23 & From \citet{Leiner2019}\\
        $f(m)$ &  $2.38 \times 10^{-3}$ & From \citet{Geller2021}\\\
        M$_\text{2,min}$  &  0.15 \Msolar & Calculated from $f(m)$ \citep{Leiner2019} \\
        M$_\text{2,orbit}$ & 0.35 \Msolar & Predicted from $P_{orb}$ \& ecc. using \citet{Rappaport1995}\\
        Age (Myr) & $\sim 300$  & Using \citet{Angus2019} gyrochronology models\\
        \hline
    \end{tabular}
    
    \caption{Stellar and Orbital Properties of WOCS 14020}
    \label{tab:WOCS14020}
\end{table}


\section{Spectral Fitting}
\label{sec:analysis}
To constrain the WD parameters, we fit the reduced and combined COS spectrum with WD atmosphere models \citep{Koester2010} using the MCMC tool \texttt{emcee} \citep{Foreman13}. An MCMC approach best captures the inherent degeneracy between $\log g$ and $T_{\rm{eff}}$ for WD atmosphere fits. The normalization of the fitted WD atmosphere scales with $r^2/d^2$, where $r$ is the WD radius and $d$ is the distance to the system. We adopt the M67 cluster distance from \citet{Stello2016} of $816 \pm 11$ pc. The radius of a WD depends on both the surface gravity and temperature, as well as the core composition of the WD. Rather than make an assumption about the core composition, we include a third ``core-picking" parameter that chooses whether to fit using a He-core composition \citep[corresponding to $\log g < 7.7$,][]{Althaus2013} or a CO-core composition \citep[corresponding to $\log g \geq 7.7$,][]{Holberg2006, Tremblay2011}. We apply flat priors on $\log g$ and $T_{\rm{eff}}$ ranging from 6.0--9.0 and 10000--18000 K, respectively, and a flat prior across the core-picking parameter. 

\begin{figure}
    \centering
    \subfigure[]{\includegraphics[width=0.8\linewidth]{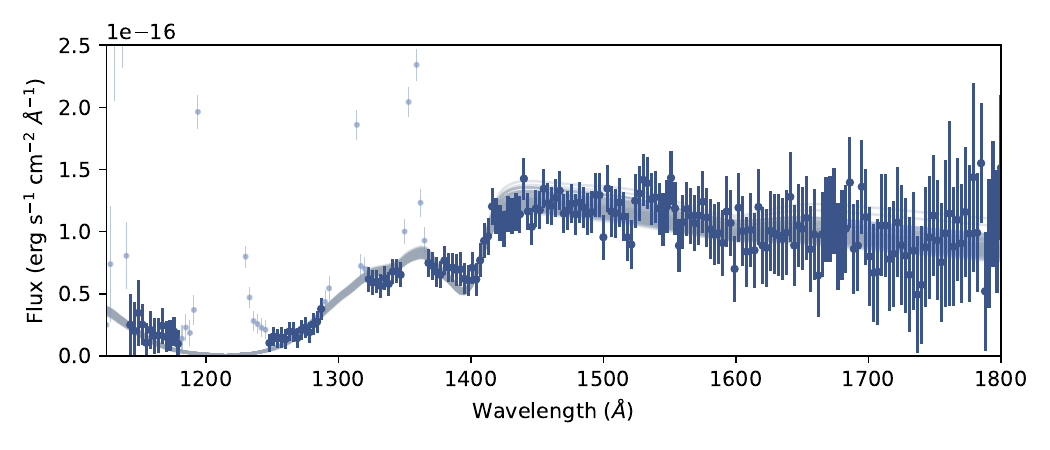}}
    \subfigure[]{\includegraphics[width=0.8\linewidth]{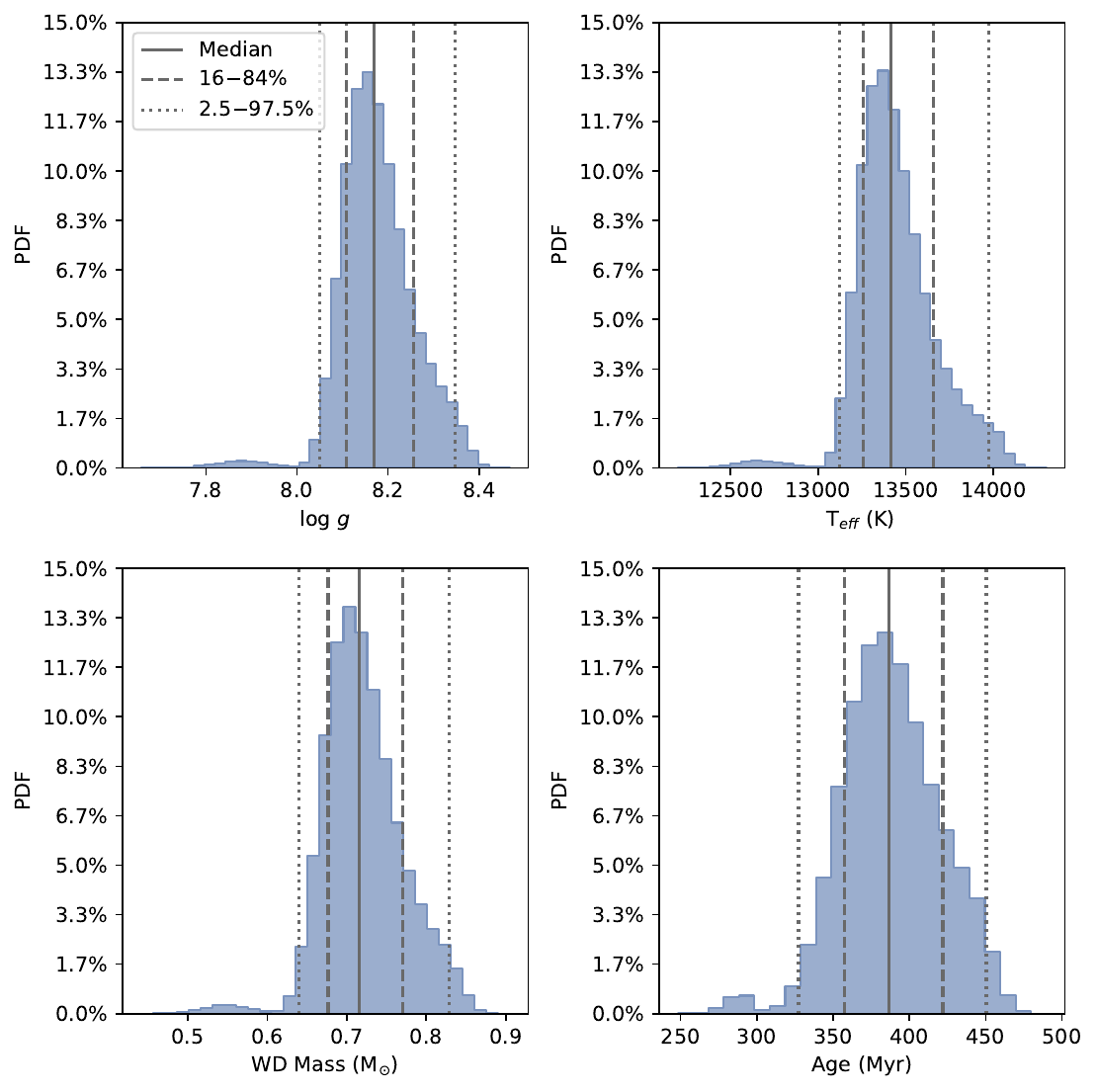}}

    \caption{(a) Binned COS spectrum for WOCS 14020, overlaid with 100 random draws from the posterior distribution from the WD atmosphere fits. The light blue data points correspond to Earth-shine emission and are not included in the fitting routine. (b) Posterior probability distributions for the WD atmosphere fits to the COS spectrum, with the median, 16--84\% (approximating 1 sigma), and 2.5--97.5\% (approximating 2 sigma) percentile values marked as indicated in the legend. The $\log g$ and $T_{\rm{eff}}$ values are fit directly and the corresponding WD mass and age are calculated using CO-core mass-radius relationships and cooling times.}
    \label{fig:spec_and_fits}
\end{figure}

 We ran 300 walkers for 30000 steps with a thinning factor of 10 and a burn-in of 1000, for $6\times10^{5}$ final samples. The resulting autocorrelation times for $\log g$, $T_{\rm{eff}}$, and the core-picking parameter are 8.6, 8.2, and 6.3, respectively, demonstrating that the model fits are stable. The model fitting posteriors are shown in Figure~\ref{fig:spec_and_fits}, with best fit values (using 16th and 84th percentiles) of $\log g = 8.17^{+0.09}_{-0.06} \textrm{ cm s}^{-2}$ and $T_{\rm{eff}} = 13400^{+240}_{-160}$ K. From the fitted $\log g$ and $T_{\rm{eff}}$ values, we calculate the corresponding core mass and cooling age ranges by interpolating standard WD mass-radius relationships \citep{Holberg2006, Tremblay2011, Althaus2013}. The resulting derived core mass and cooling age are $M_{\textrm{WD}} = 0.72^{+0.05}_{-0.04} M_{\odot}$ and $390^{+40}_{-30}$ Myr. 
 
 A He-core WD is completely eliminated as a reasonable fit to the spectrum. The \texttt{emcee} results definitively point not only to a C/O-core WD, but a \textit{surprisingly massive} C/O-core WD of approximately 0.7 \Msolar\ given its very young age. We note there is a small portion of the posterior distribution (approximately 2\%) that allows for a moderate WD mass of 0.55 \Msolar, as can be seen on the left edge of the posterior distributions in Figure~\ref{fig:spec_and_fits}, but the vast majority of the posterior is consistent with a massive WD companion.

\section{Formation Pathway}
\label{sec:formation}
The measured mass of the WD companion star, $0.72^{+0.05}_{-0.04}$ \Msolar\ is significantly larger than the predicted WD mass that would result from end-state evolution of any turnoff mass star at the age of M67 ($M_\text{WD}\sim0.55$ \Msolar). The WD age we determine is $\sim400$ Myr, and the age of M67 is generally found to be between 3.5-4.2 Gyr \citep{Stello2016, Sarajedini2009, Barnes2016}. Based on this age range, the mass transfer would have occurred when a typical giant star in M67 would have been $\lesssim1.6$ \Msolar. According to the \texttt{COSMIC} (\citealt{Breivik2020}, a rapid population synthesis code based on pre-calculated \texttt{BSE} stellar evolutionary tracks), this would yield a final WD mass of $<0.57$ \Msolar. Therefore, our measured WD mass is significantly more massive than a WD that could have been produced by a typical single star in M67. 

Given standard single-star stellar evolution, a massive WD of 0.72 ~\Msolar\ requires a progenitor with a mass close to 3~\Msolar. Such a massive star would have a main sequence lifetime of $\sim800$ Myr, and therefore 0.72 ~\Msolar\ WDs in M67 has an expected cooling age of $\sim3$ Gyr. The observed cooling age of $\sim400$ Myr for the WD in WOCS 14020 necessitates that a $3$~\Msolar\ star was created in the cluster quite recently, likely through a binary merger or interaction. We note that \citet{Williams2018} (see also \citealt{Canton2021}) detect a single WD with a similar mass in M67, and also infer that it descended from a stellar merger. Given the 400 Myr cooling age of the WD, we can infer the merger occurred at least 400 Myr ago, and perhaps as much as $\sim 1$ Gyr ago if enough new hydrogen was mixed into the core of the merger product that the $\sim3.0$ \Msolar\ merger product was significantly rejuvenated. Therefore, at the time of merger, the main-sequence turnoff in the cluster would have been 1.5-1.6 \Msolar, and a merger of two near-turnoff stars would yield a star of the required mass.

\begin{figure}
    \centering
    \subfigure[]{\includegraphics[width= 0.5\linewidth]{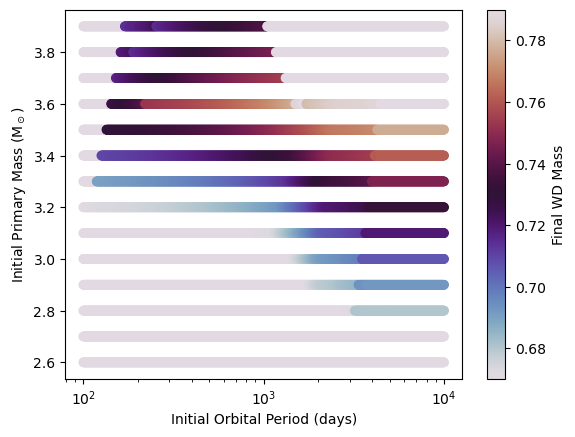}}
   \subfigure[$\alpha= 1.0$]{ \includegraphics[width= 0.49\linewidth]{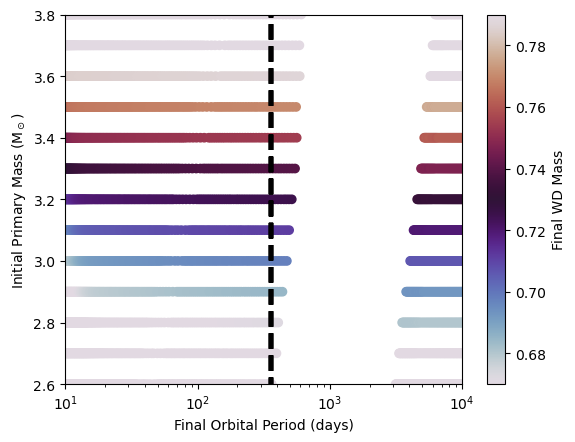}}
   \subfigure[$\alpha= 0.8$]{ \includegraphics[width= 0.49\linewidth]{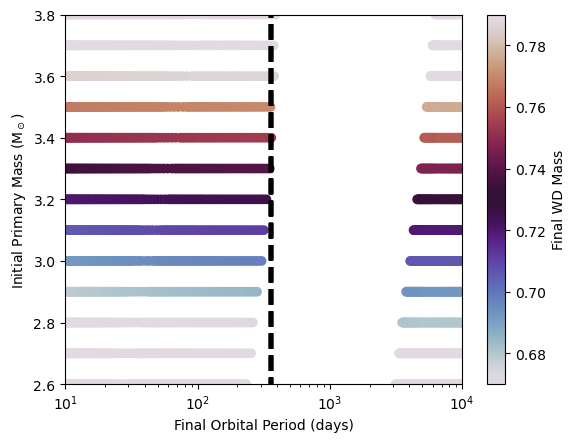}}
    \caption{(top) \texttt{COSMIC} model grid showing the initial primary mass and initial orbital period of our post-merger binary. Colors indicates the mass of the C/O WD that emerges after the final CE and emphasize final WD masses that fall within our mass confidence interval (0.68-0.78 \Msolar). (bottom) The progenitor masses and orbital periods of the binary system \textit{after} evolution through mass-transfer or a CE. On the left we show models using $\alpha= 1.0$ and on the right we show models using $\alpha= 0.8$. The 359-day orbital period of WOCS 14020 is indicated with the vertical dashed line.}
    \label{fig:models}
\end{figure}

In Figure~\ref{fig:models}, we show \texttt{COSMIC} model grids of potential progenitor binary systems. We use default \texttt{COSMIC} parameters on a grid of solar metallicity main sequence binaries with initially circular orbits between 100 and 10000 days. We vary the initial primary mass from 2.4 to 3.9 \Msolar, and use a fixed secondary mass of 1.0 \Msolar. The colors indicate the final remnant WD mass, with colors chosen to emphasize the regions where the WD mass falls within the 16th--84th percentile region of our WD mass determination, what we will refer to as our confidence interval. According to panel (a), the progenitor of this WD must be larger than 2.8 \Msolar\ to produce a WD with a mass within our confidence interval. Progenitors with masses approaching $4.0$ \Msolar\ are also possible if they are in binaries with $P_\text{orb} \lesssim 10^3$ days, where the core growth will be truncated by the onset of mass transfer. However, given that the turnoff mass at the time of formation was less than 1.6 \Msolar, the maximum mass of a star produced via a merger of two stars in the cluster would be 3.2 \Msolar. A star more massive than 3.2 \Msolar\ would require at least 3 stars to merge or interact in order to form. While this is possible, production of stars more massive than twice the turnoff via multiple stellar interactions is expected to happen  infrequently in an open cluster like M67, while mergers of main-sequence stars are common (e.g. \citealt{Hurley2005}).  Therefore, we suggest the most likely progenitor of this WD was a $2.8-3.2$ \Msolar\ star, approximately twice the turnoff mass at the time of the merger. We note that this would require a highly efficient merger with $\lesssim 0.4$ \Msolar\ of mass loss. Mass transfer efficiency is a major outstanding question in binary evolution, but high efficiencies are often favored for main-sequence mass-transfer and mergers (e.g. \citealt{Hurley2002, Sills2005,  DeMink2007, Chen2008b, Henneco2024}). Within this mass range, the progenitor binary would have had an initial orbital period of a few thousand days or more to produce a WD of the observed mass. 

In panels (b) and (c) of Figure~\ref{fig:models}, we show a similar plot to (a), but plotting the final orbital periods of the binary grid after mass transfer.  We show the observed orbital period of the BL-WD binary with a dashed black line. These models indicate that mass-transfer in this system would have been unstable, resulting in a common envelope (CE) that shrinks the orbital period. Initial orbital periods of a few thousand days will shrink to periods of a few hundred days or less, depending on the efficiency of the CE ejection, matching the observed 359-day period of WOCS 14020. We use \texttt{COSMIC} default parameters, except we vary the CE efficiency $\alpha$. This $\alpha$ parameter describes the fraction of the orbital energy that can be used to unbind the giant envelope (see \citealt{Hurley2002} and \citealt{Breivik2020} for a complete description of the CE treatment in BSE/COSMIC). The CE $\lambda$ parameter, which relates to the stellar envelope binding energy, is set using the default \texttt{COSMIC} option of adopting the value from \citet{Claeys2014}. This CE prescription can produce a binary with the observed orbital period of WOCS 14020 following an episode of CE evolution, provided the CE ejection efficiency is high ($\alpha > 0.8$.). We note, however, that our models assume all binaries start in circular orbits, and stay circularized as they evolve. WOCS 14020 is currently in a moderately eccentric orbit. This eccentricity is difficult to reproduce in models because it is not theoretically understood, which we discuss further in Section~\ref{sec:discussion:ecc}. 

Based on this modeling, the observed stellar and orbital properties of the BL-WD binary, and the additional constraints imposed by the star's membership in M67, we propose this system likely formed via the merger of an inner binary in a hierarchical triple, followed by mass transfer from the merger remnant onto the wide tertiary companion resulting in a CE. This triple may have been primordial or may have formed dynamically. In more detail:  

\begin{enumerate}
\item {This system began as a hierarchical triple. The inner binary consisted of two main-sequence stars with near equal masses of M $\sim 1.4-1.6$ \Msolar\ in a short-period binary system. The system had a main-sequence tertiary companion with M$\sim 1.0$ \Msolar\ in a wide orbit of several thousand days. }

\item {Between $\sim400$ Myr and $\sim1$ Gyr ago, the inner binary underwent a merger, perhaps induced by stellar evolution as one of the components began to evolve off the main sequence, magnetic braking, or due to Kozai-Lidov cycles \citep{Kozai1962, Lidov1962} induced by the tertiary companion (e.g. \citealt{Perets2009, Naoz2014}). Assuming a high efficiency, the merger remnant was $\sim3.0$ \Msolar\ and retained the initial tertiary as a binary companion with an orbital period of $P_\text{orb} \sim 3500$ days.}

\item{About 400 Myr ago, the $\sim 3.0 $ \Msolar\ merger remnant evolved into an AGB star. Near the tip of the AGB, wind mass transfer began and the original tertiary companion accreted a small amount of material via wind. This spun up the rotation rate of the accretor, yielding the rapid rotation we see in the BL today.}

\item{Soon a CE was triggered and the binary inspiraled.  A high common envelope efficiency resulted in the common envelope being quickly lost from the system. Its initial period of $\sim3500 $ days shrank to the currently observed period of 359 days. }

\item{
The WD began to cool and the BL spun down via magnetic braking. }

\item{We now observe the BL-WD binary as it is today: a 1.05 \Msolar\ BL primary with  a 0.72 \Msolar\ WD secondary in a binary with $P_\text{orb}= 359$ days. }

\end{enumerate}

In summary, our observed WD constraints plus COSMIC models point to this BL-WD system likely forming through the merger of the inner binary in a hierarchical triple system, followed by a common envelope event when the merger product evolves to be an AGB star. This places WOCS 14020 as an excellent system for future dynamical modeling efforts, including hydrodynamical modeling of the merger, to illuminate this triple  evolutionary pathway in more detail.

\section{Discussion}
\label{sec:discussion}

We have argued above that WOCS 14020 is a remarkable example of the complexity of evolution in a triple system. We propose that its evolution has included a merger of an inner binary in a hierarchical triple, followed by a common envelope between the merger product and the original tertiary. Here we discuss several  implications of this case-study in triple star evolution.

\subsection{Common Envelope Evolution}
The current orbital period of WOCS 14020 indicates that the system must have evolved through a CE, as the separation is much too small to host a giant star with a 0.72 \Msolar\ core.  Given a progenitor donor star of $2.8-3.2$ \Msolar\ and the current WD mass of 0.72 \Msolar, the orbital period at the onset of the CE must have been a few thousand days (Figure~\ref{fig:models}, b and c). The CE ejection efficiency must have been quite high to result in the observed amount of orbital decay. \texttt{COSMIC} models require a value of the CE $\alpha$ of 0.8--1.0 to reproduce the observed post-CE orbital period of this system.  This CE efficiency is a large uncertainty in binary evolution, with some studies arguing that more inefficient envelope ejections with $\alpha \sim 0.2-0.3$ better match observed the characteristics of the post-CE MS-WD population (see, for example \citealt{Zorotovic2010, Toonen2013, Camacho2014} or double WD systems \citealt{Scherbak2023MNRAS}), while others require $\alpha \sim 1.0$ or even larger \citep{DeMarco2011,SunM2018}. Most previous investigations have looked at post-CE binaries with very short orbital periods of a few days or WDs with low-mass M-dwarf companions to constrain $\alpha$ (e.g. \citealt{Zorotovic2010, Scherbak2023MNRAS}), generally finding a much lower efficiency of $\sim 0.2 \lesssim \alpha \lesssim 0.4$ is required. Our observations offer a rare constraint on outcomes of CE that involve solar-type companions and result in orbital periods of hundreds of days. Some similar systems have recently been discovered that also favor efficient ejection of CEs initiated during the donor's AGB phase \citep{Belloni2024, Yamaguchi2024a}, as does the recent discovery of a large population of MS-WD binaries in the field with orbital periods in the range of $10^2-10^3$ days \citep{Yamaguchi2024b}. Our observation and analysis also favor this efficient envelope ejection. 

\subsection{Orbital Eccentricity}
\label{sec:discussion:ecc}
Notably, WOCS 14020 has a non-zero eccentricity of e = 0.23. Outcomes of stable mass transfer and CE evolution have often been assumed to be circular. However, it is common for blue stragglers, BLs, and related post-mass-transfer binaries such as binary post-AGB, s-processed enhanced stars, and main sequence-WD binaries to have non-zero eccentricities (e.g. \citealt{Mathieu2009, Leiner2019, Oomen2018,  Escorza2019, Shahaf2024, Yamaguchi2024b}). The reasons for this are still uncertain. Some work has proposed that eccentricity pumping during stable mass transfer can occur (\citealt{Sepinsky2007,Sepinsky2009,Sepinsky2010}; Rocha et al. 2024, in prep), which may explain some of these eccentric post-mass-transfer binaries, but WOCS 14020 does not appear to have evolved through stable mass transfer. CE events are generally assumed to result in circularized binary systems, but a growing body of work also questions this assumption. Accretion from a circumbinary disk is a potential mechanism to excite eccentricities in binaries \citep{Dermine2013}, including post-CE binaries with circumbinary disks \citep{Kashi2011,Siwek2023, Valli2024, Wei2024}. Alternatively, the binary may not actually circularize prior to the 
CE as is often assumed, and some eccentricity may be preserved after the CE (e.g. \citealt{Bonavic2008, Prust2019}). 

\subsection{Mass Accretion and Spin Up}
The rapid rotation of the BL argues for some accretion during the system's evolution. It is often assumed that no accretion occurs on to the secondary star that evolves through CE evolution, but the rapid rotation requires some accretion happened before, during, or after the CE event. \citet{Sun2024} recently showed that even a small amount of wind accretion ($< 0.1$ \Msolar) can spin up accretors to critical rotation rates. In our \texttt{COSMIC} model, the proto-BL accretes $\sim 0.1$ \Msolar\ via a wind prior to the onset of CE, and this may be enough to explain the observed rotation. \citet{Nine2024} recently also surveyed the blue stragglers and BLs in M67 for Barium enhancement, an often used indicator of mass transfer from an AGB companion. They did not find WOCS 14020 to be significantly barium enhanced relative to the main sequence population. 

This finding is broadly consistent with our model, as the amount of accretion is minimal and the donor star's AGB phase is terminated early due to the onset of the CE. Since barium is produced during the late-stage thermal pulses, the barium yield would likely have been much lower than predicted for a 3.0 \Msolar\ star evolving in isolation. While we note that only minimal accretion is needed to produce the rapid rotation and observed color-magnitude diagram position of the BL, our observations do not rule out larger accretion amounts, and other accretion mechanisms besides a wind prior to CE onset may also be possible. For example, there could be accretion from a circumbinary disk \citep{Lai2023}) that forms after the CE \citep{Kashi2011}, which could transfer mass to the BL, spin it up, and lead to the observed orbital eccentricity as noted in the previous section.

\subsection{The Importance of Triples}
Given what we know about stellar multiplicity and binary statistics, evolutionary paths similar to WOCS 14020 may be fairly common. Around 10\% of solar-type stars are found in triple systems \citep{Raghavan2010}, and a configuration with a short-period near-equal mass inner binary and a wider tertiary is common among these systems. \citet{Moe2017} find an excess of ``twin" binaries among short period ($P < 100$ day) solar-like binaries such that $\sim 30\%$ of short-period binaries have near equal mass ratios ($0.95 <\frac{M_2}{M_1} < 1.0$). A large fraction of close binaries have also been found to have wide tertiary companions; \citet{Tokovinin2006} find the overall triple fraction among close spectroscopic binaries ($P_\mathrm{orb} < 30$ days) is $63 \pm 5\%$, and that rises to 96\% among the closest binaries with $P_\mathrm{orb} < 3$ days. Therefore, mergers of inner binaries in triples should often form stars that are nearly twice the turnoff mass of a cluster, and these merger products will often have  companions they will interact with later in their evolution. 

Growing observational evidence also points to the general importance of triples in stellar evolution. \citet{Heintz2022} analyze the \textit{Gaia} sample of wide double WD binaries, which reveals a large fraction ($\sim 20$\%) of systems include a WD that resulted from a merger, and thus originated in a triple system. \citet{Shariat2024} also argue that in $\sim40$\% of wide double WD binaries, the more massive WD is a merger product, and thus these systems are descended from triples. Further, they find that 20-25\% of blue stragglers may form from evolution in triples. 

In M67, \citet{Leiner2016} detected an overmassive giant with a binary companion, S1237. This system consists of a red giant primary that is an asteroseismic outlier for the cluster. Asteroseismic analysis yields a mass of $2.9 \pm 0.2$ \Msolar, more than twice the turnoff mass of  M67, and indicates the star is likely a core helium burning giant. This overmassive giant has a binary companion in a 697.8 day orbit, which seems to be located near the cluster turnoff or perhaps in the blue straggler region. One likely formation scenario for S1237 is quite similar to WOCS 14020: the overmassive giant likely resulted from the merger of a close, near-equal-mass inner binary in a hierachical triple. The current binary companion would then have previously been a wide tertiary. The giant in S1237 is expected to form a WD in the near future ($\sim 100$ Myr), at which point the system should be observed as a blue straggler- or BL-WD binary.  Another blue straggler system in M67, S1082, likely is currently a multiple system containing two blue stragglers, and may have formed from multiple mergers or collisions given the system's large combined mass \citep{Sandquist2003, Leigh2011}. Descendants of triple systems are thus not rare amongst the known post-mass-transfer population of M67; at least one appears to be produced every few hundred Myr. 

Theoretical work has also highlighted the potential importance of creating blue stragglers, double WD binaries, barium stars, and other post-interaction systems via triples (e.g. \citealt{Perets2009, Toonen2022, Gao2023}).  This reflects a growing theoretical consensus that considering triple systems is vital to understanding the full breadth of stellar evolution pathways and outcomes. WOCS 14020 is among the best observationally-characterized examples of a post-interaction triple system to date, and thus is a much needed test case for future models of evolution in triple systems.

\subsection{Future Evolution of WOCS 14020}
Intriguingly, WOCS 14020 could be a potential Type Ia supernova progenitor via a sub-Chandrasekhar double detonation \citep{Liu2023}, or the progenitor of a calcium-rich transient \citep{Kasliwal2012, JacobsonGalan2021, MoranFraile2024}. Based on past detailed case studies of open cluster blue stragglers forming through mass transfer \citep{Sun2021, Sun2023}, in approximately 5 Gyr, the BL will evolve into a giant star and begin to interact with the C/O WD companion. It is improbable that WOCS 14020 will be disrupted by a dynamical encounter within the next 5 Gyr \citep{Leigh2011}, so likely the system will evolve towards a double WD system of some kind. Most likely, a CE will occur, leaving behind a close double WD binary consisting of the 0.72~ \Msolar\ C/O WD and a 0.38 \Msolar\ He WD that is the remnant of the current BL. Depending on the CE efficiency, this double WD binary may be close enough to interact, with the larger He WD accreting onto the C/O WD. The outcomes of such sub-Chandrasekhar mass mergers and interactions are still not well understood, but recent models suggest a variety of interesting transients may result. 

In a detailed binary evolution study, \citet{Wong2023} simulated a close He WD transferring mass onto a C/O WD and discussed the necessary conditions for triggering a type Ia supernova event. One of the requirements is to have a high-entropy He WD, which could be formed from a CE event. In general, our system configuration aligns well with \citet{Wong2023} for achieving a successful explosion event. Furthermore, it has recently been suggested that C/O WD with M $< 1.0$ \Msolar\ can undergo a detonation by accreting stably or unstably from a He WD companion \citep{Shen2024}, resulting in a normal Type Ia supernova. Some Type Ia supernova have been observed with similar C/O WD progenitor masses (see \citealt{De2019, LiuChang2023}). Mergers between C/O and He WD have also been suggested as potential sources of calcium-rich supernovae (e.g. \citealt{MoranFraile2024}). The evolutionary history of WOCS 14020 underscores that triple evolution may be a source of late-time transients, including double detonation Type Ia supernova and calcium-rich transients.

\section{Summary and Conclusion}
\label{sec:summary}
WOCS 14020 is a BL-WD binary in the open cluster M67. It has a-359-day orbital period with a moderate eccentricity of e= 0.23. We fit the Lyman-$\alpha$ region of a COS far-UV spectrum of a WD star in the BL-WD binary system WOCS 14020 in the 4 Gyr open cluster M67. From this fit we determine $\log g = 8.17^{+0.09}_{-0.06} \textrm{ cm s}^{-2}$ and $T_{\rm{eff}} = 13400^{+240}_{-160}$ K for this WD. This corresponds to a C/O WD with $M_{\textrm{WD}} = 0.72^{+0.05}_{-0.04} M_{\odot}$ and a cooling age of $390^{+40}_{-30}$ Myr. The mass is significantly larger than expected for a typical WD of this age in this cluster, and we argue that the progenitor star was a $2.8-3.2$ \Msolar\ merger product. We suggest a formation scenario in which this BL-WD binary is descended from a hierarchical triple system in the cluster, which undergoes first a merger of the inner binary, followed by mass transfer onto the outer tertiary that ends with a CE that shrinks the orbit from $\sim3500$ days to its currently observed 359 day period. We note that:
\begin{itemize}
\item The rapid rotation of the BL indicates it accreted at least a small amount of material during the interaction, possibly via a wind prior to CE and/or accretion from a post-CE disk. 
\item The non-circular orbit indicates eccentricity is either somehow  maintained through the CE evolution or excited after the CE. 
\item the current orbital period of 359 days indicates an efficient envelope ejection ($\alpha \gtrsim 0.8$). 
\end{itemize}
Additional hydrodynamical modeling of the common envelope evolution may provide further insights into the evolutionary path of this system and illuminate the cause of the features noted above.

 In the future, this system will likely undergo a second CE when the BL evolves into a red giant and overflows its Roche lobe. Ultimately, this may form a close double WD binary or yield a He WD-C/O WD merger. The story of WOCS 14020 thus underscores the complexities of evolution in triple star systems, and highlights that triple evolution is an important channel creating blue stragglers, double WD binaries, explosive transients, and other important astrophysical objects that result from stellar interactions. WOCS 14020 is the first blue lurker system for which an evolutionary path has been illuminated in detail, a result made possible because the stellar and orbital parameters of both binary components have been exceptionally well-determined by observations. This case study offers rare and intriguing insights about the outcome of interactions in a triple star system.

\section*{Acknowledgments}
This paper is based on observations made with the NASA/ESA \textit{Hubble Space Telescope}, obtained at the Space Telescope Science Institute, which is operated by the Association of Universities for Research in Astronomy, Inc., under NASA contract NAS 5-26555. These observations are associated with program \#17134 and supported by NASA HST-GO-17134. This work has made use of data from the European Space Agency (ESA) mission
{\it Gaia} (\url{https://www.cosmos.esa.int/gaia}), processed by the {\it Gaia}
Data Processing and Analysis Consortium (DPAC,
\url{https://www.cosmos.esa.int/web/gaia/dpac/consortium}). Funding for the DPAC
has been provided by national institutions, in particular the institutions
participating in the {\it Gaia} Multilateral Agreement. E.L. is additionally supported by funding from Illinois Institute of Technology. M.S. acknowledges the support from the GBMF8477 grant (PI: Vicky Kalogera) and thanks Chang Liu and Sunny Wong for discussions on double detonation type Ia supernovae. R.M acknowledges funding from the Wisconsin Alumni Research Foundation. A.S. is supported by the Natural Sciences and Engineering Research Council of Canada. 

\software{\texttt{COSMIC} \citep{Breivik2020}, \texttt{emcee} \citep{Foreman13}
\texttt{numpy} \citep{2020NumPy-Array}, \texttt{Matplotlib} \citep{Hunter2007},\texttt{astropy}\citep{Astropy2013}}
\\
\\
All the {\it HST} data used in this paper can be found in MAST: http://dx.doi.org/10.17909/twcy-vq55

\bibliographystyle{apj}
\bibliography{BSScitations}
\end{document}